\documentclass[preprint,12pt]{elsarticle}

\usepackage{amssymb}
\usepackage{amsmath}

\usepackage{bbm}
\usepackage{subcaption}
\usepackage{xcolor}
\usepackage{algorithm}
\usepackage{algpseudocode}
\usepackage{multirow}
\usepackage{booktabs}
\usepackage{xurl}
\usepackage[capitalise]{cleveref}
\crefformat{appendix}{#2#1#3}

\journal{Sustainable Energy, Grids and Networks (SEGAN)}

\begin{document}

\begin{frontmatter}



\title{Probabilistic Load Forecasting of Distribution Power Systems based on Empirical Copulas}


\author[1]{Pål Forr Austnes}
\author[2]{Celia García‐Pareja}
\author[2]{Fabio Nobile}
\author[1]{Mario Paolone}

\affiliation[1]{organization={Ecole Polytechnique Fédérale de Lausanne, EPFL},
            addressline={Distributed Electrical Systems Laboratory (DESL)}, 
            city={Lausanne},
            postcode={1015}, 
            state={Vaud},
            country={Switzerland}}
\affiliation[2]{organization={Ecole Polytechnique Fédérale de Lausanne, EPFL},
            addressline={Scientific Computing and Uncertainty Quantification CSQI}, 
            city={Lausanne},
            postcode={1015}, 
            state={Vaud},
            country={Switzerland}}

\begin{abstract}
Accurate and reliable electricity load forecasts are becoming increasingly important as the share of intermittent resources in the system increases. \emph{Distribution System Operators} (DSOs) are called to accurately forecast their production and consumption to place optimal bids in the day-ahead market. Violations of their dispatch-plan requires activation of reserve-power which has a direct cost for the DSO, and also necessitates available reserve-capacity. Forecasts must account for the volatility of weather-parameters that impacts both the production and consumption of electricity. If DSO-loads are small or lower-granularity forecasts are needed, parametric statistical methods may fail to provide reliable performance since they rely on a priori statistical distributions of the variables to forecast. In this paper, we introduce a \emph{Probabilistic Load Forecast} (PLF) method based on Empirical Copulas (ECs). The model is data-driven, does not need a priori assumption on parametric distribution for variables, nor the dependence structure (copula). It employs a kernel density estimate of the underlying distribution using beta kernels that have bounded support on the unit hypercube. The method naturally supports variables with widely different distributions, such as weather data (including forecasted ones) and historic electricity consumption, and produces a conditional probability distribution for every time step in the forecast, which allows inferring the quantiles of interest. The proposed non-parametric approach differs significantly from previous forecasting methods based on copulas, which typically uses copulas to model hierarchical dependence. Our approach is highly flexible and can produce meaningful forecasts even at very low aggregated levels (e.g. neighborhoods). The bandwidth of the beta kernel density estimators is optimized using \emph{Integrated Square Error} (ISE) and such optimization can be performed online (i.e. without knowing the realization). We also investigate rule-of-thumb and \emph{Quantile Loss} (QR) as objectives for the bandwidth-optimization. We present results from an open dataset and showcase the strength of the model with respect to \emph{Quantile Regression} (QR) using standard probabilistic evaluation metrics. 
\end{abstract}

\begin{graphicalabstract}
\end{graphicalabstract}

\begin{highlights}
\item Data-driven probabilistic electricity load forecasts at different aggregation levels based on empirical copulas
\item The proposed method allows for capturing nonlinear relations between variables and does not make any assumption on probabilistic distribution
\item The proposed method can easily be extended to include any type of exogenous variables
\end{highlights}


\end{frontmatter}



\section{Introduction}
Forecasts on a low level of aggregation and hierarchical forecasts help system operation, may facilitate the quantification of demand-flexibility and have been enabled by the widespread introduction of smart meters. Point-forecasts have been thoroughly studied in the literature \cite{Nti2020}. Traditional forecasting methods, such as multivariate regression and auto-regressive models, assume a parametric model of the residuals and the uncertainty quantification is therefore limited to the degrees of freedom of the parametric model. In recent years, several approaches using \emph{Probabilistic Load Forecast} (PLF) techniques have also been studied (e.g. \cite{book-pinson,pinson_estimation_nodate, NESPOLI2020106755}). In addition to the expected value, PLFs also provides prediction or confidence intervals and can, therefore, quantify the forecast uncertainty \cite{HONG2016914}. Another method widely studied is the \emph{Quantile Regression} (QR) and its extension, the \emph{Quantile Regression Averaging} (QRA). QR is analogous to linear regression, but instead of estimating the conditional mean it estimates quantiles. However, QR might experience quantile crossing, a phenomenon where the monotonicity of the predicted quantile function is violated, resulting in non-coherent forecasts \cite{doi:10.1080/00031305.1997.10473959}. QRA requires an initial point-forecast and estimates the prediction intervals from the observed residuals between the point-forecast and the realization \cite{hong_global_2019,9218967}. QRA is powerful in that it can combine forecasts from forecasting models and experts to enhance the performance of individual models. However, it requires larger datasets, as parts of the dataset must be reserved for fitting the individual models.

In this paper we introduce a PLF method based on empirical copulas. As known, a copula is a multivariate distribution function with every marginal distribution being a standard uniform distribution. Sklar's theorem states that any multivariate distribution can be represented, upon a suitable transformation, by uniform marginals and a copula that describes the dependence structure between the variables \cite{Sklar1959FonctionsDR}. Empirical copulas do not make any assumption on the underlying probability distribution, nor the dependence structure (e.g. linearity) between variables, which allows data-driven forecasts on any aggregation-level

The paper is divided as follows: in \cref{section:related-work} we present related works and motivate the need for accurate PLF. In \cref{sec:model,sec:estimation-procedure} we present the model and the estimation procedure. In \cref{sec:case-study} we present the case study and apply the developed method to a public dataset and evaluate its performance against QR. Finally, in \cref{section:discussion-conclusion} we discuss the results and conclude.


\section{Related work and motivation} \label{section:related-work}
\subsection{History}
Research on \emph{Electricity Load Forecasting} (ELF) has seen a renaissance in the later years, driven by more powerful computational resources and the large introduction of stochastic electricity production. However, the research topic goes back several decades. In the early beginnings of country-wide vertically integrated utilities, the main challenges were related to accurately modeling expansions of production and transmission lines capacity \cite{5060241}. The electricity-boom of the 1960s, with very large growth of electricity-demand, followed by the 1970s energy-crisis led to large stresses on the power grid. In the 1970s and 1980s, the focus of research shifted towards economic dispatch modeling and peak demand forecasts \cite{osti_5195319}, together with multi-price schemes to stimulate off-peak demand. The increased lead-time for constructing new plants and the associated cost-increases were also reasons for increased interest in ELF \cite{emf1979}. 

Until the 1990s, the electricity-supply was controlled by monopolies. However, with the aim of increasing competition, the monopolies of utilities were broken and liberalized electricity-markets introduced. This move, primarily driven by legislators in the European Union and in the United States, sparked renewed interest in ELF, and also in electricity price forecasting. In combination with the adoption of machine learning concepts, such as neural networks, the complexity of the forecasting models increased significantly. A survey of practical implementations of load forecasting techniques by utilities in 1992 showed that the complexity of methods varied greatly \cite{cigre_wg}. The most popular methods were variants of multiple linear regression, Box-Jenkins and exponential smoothing. However, the survey respondents point out unacceptable prediction errors, lack of weather-parameters, necessity of online methods and the forecasting of special days as the main reasons for insufficient performance of the forecasting models. 

In the early 2000s and 2010s, the large introduction of stochastic distributed electricity production pushed the need for better forecasting tools. The adoption of probabilistic forecasts seemed a suitable approach to allow quantification of the uncertainties of forecasts and have been successfully applied to wind \cite{pinson_estimation_nodate} and solar energy \cite{LI202068} production. In a similar fashion, it has also been applied for ELF \cite{NESPOLI2020106755}. 

Increased stochastic distributed electricity generation has increased the need for balancing reserves (e.g. \cite{elcom2021}). Violations of the day-ahead dispatch-plan activates reserves by the \emph{Transmission System Operator} (TSO), resulting in additional costs for \emph{Distribution System Operators} DSOs and balancing groups. \emph{Active Distribution Networks} (ADNs) give the flexibility to actively dispatch the local grid and optimize resource-management. Accurate and reliable forecasts help minimize the needed reserves and help the integration of renewable electricity-production, ultimately lowering the overall cost of the system. Furthermore, probabilistic forecasts allow for the assessment of the expected uncertainty and optimal bidding strategies in the electricity markets. Combining probabilistic forecasts with dispatchable resources such as \emph{Battery Energy Storage Systems} (BESSs), through stochastic optimization-routines has been shown to track the dispatch-plan of distribution-grids to a very high accuracy (e.g. \cite{9925205}). Forecasts at a low level of aggregation can also enhance overall forecasts, by better capturing patterns of individual consumers \cite{hong_global_2019}. 

\subsection{Literature review}
Forecasts are usually divided into 4 categories: very-short term (seconds to minutes), short-term (hours to a few days), medium-term (weeks to months) and long-term (5 years to several decades). The focus of this contribution is on short-term forecasting. 
PLF-methods such as \emph{Multivariate Normal Distributions} (MNDs) and QR assume a parametric distribution (MND) or linear dependence-structure (MND and QR), and may not be suitable at very low aggregation-levels where the dynamics are highly non-linear and variables are not necessarily (jointly) normally distributed. Several novel methods based on deep learning have been developed to address this issue. \cite{Wen_2022} uses an approach based on normalizing flows, a generative framework that allows learning a mapping between simple (e.g. Gaussian) and complex distributions. In \cite{chen2018modelfree} renewable production scenarios are generated from historical data using deep neural networks. The scenarios can be generated conditioned on specific weather events such as wind and solar irradiance. However, these black-box models lack explainability and often require substantial computational effort to be effective.

Copulas have been successfully applied to model complex dependence structures in other domains, such as modeling financial returns and hydrology, \cite{doi:https://doi.org/10.1002/9781118673331.fmatter, https://doi.org/10.1002/wat2.1579}. Copulas have also been applied to power systems. In \cite{4703184}, copulas were introduced to model stochastic dependence between variables for power systems uncertainty analysis. The authors explain that both marginal distributions and dependence cannot always be modeled with Gaussian random variables and copulas are suitable for modeling complex multivariate distributions. \cite{pmlr-v70-taieb17a} uses copulas to model the dependence structure between time series at different aggregation levels to produce coherent probabilistic forecasts for aggregate loads. In \cite{8736321}, authors propose using a combination of QR and empirical copulas to produce coherent hierarchical probabilistic forecasts, but uses multiple linear regression to forecast the individual time series.

In \cite{doi:10.1080/01621459.2020.1736081} the empirical copula was used to model the hierarchical dependence structure between households equipped with smart meters. However, the individual forecasts were performed using a kernel density estimator and not based on copulas. 

\subsection{Contribution of this paper}
Our approach differs significantly from previous works by using empirical copulas to independently model every time series in the hierarchy. While other works focus on modeling hierarchical dependence, this work proposes a model that uses historical power measurements and exogenous variables to predict individual time series. The individually forecasted time series can thereafter be aggregated into coherent forecasts using the methodologies in \cite{pmlr-v70-taieb17a, doi:10.1080/01621459.2020.1736081}. 

In summary, previous methods either assume linearity between variables and parametric description of the uncertainty to forecast individual time series, or lack explainability by being black-box models. Our model provides the following benefits:
\begin{itemize}
    \item Does not make assumptions on the parametric distribution of variables.
    \item Does not make assumptions on linear relationships between variables.
    \item The proposed framework is data-driven and can provide meaningful forecasts at different aggregation-levels.
\end{itemize}

\section{Model} \label{sec:model}
\subsection{Hypothesis} \label{subsec:hypothesis}
Our modeling setup relies on three fundamental observations, supported by previous research \cite{Nowotarski2015, WANG2016585} and are often coined 
as \emph{Auto-Regressive with eXogenous variables} (ARX) models \cite{Nowotarski2015}:

\begin{itemize}
\setlength\itemsep{1em}
 \item
Electricity demand display serial dependence with previous realizations 
 \item 
 Electricity demand at specific timestamps in similar day-types are similar too
 \item 
 Electricity demand have dependence with meteorological variables
\end{itemize}

\subsection{Problem statement}

Given time series $\{p_t\}_{t}$ and $\{\theta_t\}_{t}$ of observed values of power and temperature, and denoting $T$ the fundamental period of the time-series, in our case one day (since the sampling rate is 15-minutes, T=96), we denote by $p_j^i=p_{j+Ti}$ the value of the power at the $j$-th timestamp of the $i$-th day of the series (similarly for the temperature $\theta_j^i$). Our goal is to forecast $k\in \{1,\ldots, H\}$ steps of electricity demand for the current day, say $n$, using historical values of previous days up to $m<n$. For this, we first need to build $H$ pdfs $\hat{c}_h^{k}, \; k=1,\ldots,H$, ($h$ denoting some discretization parameter) each using the data matrix 
\begin{equation} \label{eq:data-matrix}
    X_k = \begin{pmatrix}  P_k & T_k \end{pmatrix},
\end{equation}
with:

\begin{gather}
 P_k
 =
 \begin{pmatrix}
        r(p_j^{1})       & r(p_{j-a_1}^{1}) & \dots & r(p_{j-a_{\lambda-1}}^{1})  \\
        r(p_j^{2})       & r(p_{j-a_1}^{2}) & \dots & r(p_{j-a_{\lambda-1}}^{2}) \\
        \vdots & \vdots &  \ddots & \vdots\\
        r(p_j^{m})       & r(p_{j-a_1}^{m}) & \dots & r(p_{j-a_{\lambda-1}}^{m})
    \end{pmatrix},
\end{gather}

\begin{gather}
 T_k
 =
 \begin{pmatrix}
        r(\theta_{j-b_1}^{1}) & \dots & r(\theta_{j-b_\gamma}^{1})\\
        r(\theta_{j-b_1}^{2}) & \dots & r(\theta_{j-b_\gamma}^{2})\\
        \vdots & \ddots & \vdots \\
        r(\theta_{j-b_1}^{m}) & \dots & r(\theta_{j-b_\gamma}^{m})
    \end{pmatrix},
\end{gather}
\noindent
where $\{a_1,\ldots,a_{\lambda-1}\}\in\Lambda$ are the lags in power demand and $\{b_1,\ldots,b_\gamma\} \in\Gamma$ the lags in temperature, with respect to the timestamp $j$ of the day, considered for the estimation. $r(\cdot)$ is the normalized rank-transformation function, i.e.: $r(x_i)=\frac{1}{m+1}x_{(i)}$ with $x_{(\cdot)}$ representing the rank of the sample and $m$ the number of samples. Hence $r(x_i)\in \{\frac{1}{m+1},...,\frac{m}{m+1}\}$.
\newline

In the following sections we first provide a generic case formulation for estimating joint probability distributions from data. Then we discuss copula-theory and finally the conditional probability distribution that is used for the specific task of PLF. 
\subsection{Data structure}

For a single forecasting-step, we consider the following data matrix:
\[
X=\begin{pmatrix}
    r(x_{1}^{1})       & r(x_{2}^{1}) & \dots & r(x_{\lambda}^{1}) & r(x_{\lambda+1}^{1}) & \dots & r(x_{d}^{1})\\
    r(x_{1}^{2})       & r(x_{2}^{2}) & \dots & r(x_{\lambda}^{2}) & r(x_{\lambda+1}^{2}) & \dots & r(x_{d}^{2})\\
    \vdots & \vdots & \ddots & \vdots & \vdots & \ddots & \vdots\\
    r(x_{1}^{m})       & r(x_{2}^{m}) & \dots & r(x_{\lambda}^{m}) & r(x_{\lambda+1}^{m}) & \dots & r(x_{d}^{m})
\end{pmatrix},
\]
\noindent
where every column represents a variable and every row an \emph{independent and identically distributed} (iid) sample of the model. In our model, variables $[x_{1}^{\cdot},...,x_{\lambda}^{\cdot}]$ refer to values of measured power and $[x_{\lambda+1}^{\cdot},...,x_{d}^{\cdot}]$ to meteorological variables. Each sample corresponds to the measured value of a physical quantity at a specific timestamp in a day. I.e., every row depicts a different historical day. The meteorological variables can be any of interest to model electricity demand, such as temperature, solar irradiance, wind-speed, wind-direction, humidity etc. In this paper we also include forecasted weather quantities, like the temperature. Crucially, in the estimation-procedure (\cref{sec:estimation-procedure}), the value of the meteorological variables are known and provided by a weather-forecasting service. In addition, one can include lagged values of temperatures (in a similar manner as the lagged power-values) to model the inertia of heating appliances.

\subsection{Empirical copula density estimation}
The copula is a tool to model the joint dependence between random variables. As opposed to parametric copulas such as Gauss, Clayton and Gumbel, the empirical copula makes no assumption on the shape of the marginal distributions nor their dependence structure, which are the exact properties we seek. As such, it can be regarded as a special case of an empirical distribution function where variables can only take values in $[0,1]$. Considering $D$ variables, the empirical copula is defined as \cite{deheuvels_fonction_1979}:
\begin{equation}
    C\left(u_1, ... , u_D\right) = \frac{1}{m}\sum_{i=1}^m\prod_{j=1}^D\mathbf{\underline{1}}_{r(x_j^i)\leq u_j},
\end{equation}
where $\left(u_1, ... , u_D\right) \in [0,1]^D$ and $r(x_{j}^{i})$ is the rank-function applied to the i-th sample of the j-th variable. This function is discontinuous at every $u_k \in \{\frac{1}{m},...,1\}, \forall k\in \{1,...,D\}$, and therefore, it's not suitable to obtain the corresponding \emph{probability density function} (pdf) by differentiating along every marginal variable \cite{Charpentier2007TheEO}. Several smoothing techniques have been proposed, such as kernel, wavelet, k-nearest neighbors etc. \cite{CHEN1999131} proposes a kernel density estimator using the beta kernel. The beta kernel has a natural bounded support on $[0,1]$ and therefore avoids boundary-bias. The kernel shape adapts depending on the location in the domain without need to change the bandwidth. Its definition, with $D$ variables, is: 
\begin{equation} \label{eq:beta-kernel-density}
    \hat{c}_\mathbf{h}(u_1...u_D) = \frac{1}{A}\sum_{i=1}^m\prod_{j=1}^DK\left(r(x_{j}^{i}), \frac{u_j}{h_j}+1, \frac{1-u_j}{h_j}+1 \right),
\end{equation}
where 
$A=m\prod_{j=1}^Dh_j$, $m$ is the size of the sample, $\mathbf{h} \in\mathbbm{R}^D$ is the bandwidth of the kernel and 
\begin{equation*}
    K(z,\alpha,\beta) = \frac{\Gamma(\alpha+\beta)}{\Gamma(\alpha)\Gamma(\beta)}z^{\alpha-1}(1-z)^{\beta-1}
\end{equation*}
is the pdf of the beta distribution with $\Gamma$ the Gamma function, $z\in [0,1]$ and shape parameters $\alpha$ and $\beta$. In practice, we calculate the copula on a tensor-grid with $L$ points per variable.

\subsection{Conditional density estimation} \label{subsec:conditional-density-estimation}
The formulation in \cref{eq:beta-kernel-density} can be used to evaluate the joint density of a sample of rank-normalized random variables. However, its computational complexity increases exponentially with the number of dimensions and the size of the tensor-grid. In practical terms, we observe that beyond 3 variables, the estimation of \cref{eq:beta-kernel-density} becomes computationally heavy. Instead, we directly estimate the conditional density by fixing all the variables, except the forecasted one. The formulation is very similar to the one in \cref{eq:beta-kernel-density}, except we estimate the pdf of $u_1$ in $L$ points, given fixed values of $u_2,...,u_D$.

\begin{equation} \label{eq:cond-density-estimation}
    \hat{c}_\mathbf{h}(u_1|u_2...u_D) \propto \sum_{i=1}^m\prod_{j=1}^DK\left(r(x_{j}^{i}), \frac{u_j}{h_j}+1, \frac{1-u_j}{h_j}+1 \right)
\end{equation}

The proportionality-sign indicates that the resulting conditional distribution must be normalized such that $\int_0^1\hat{c}_\mathbf{h}(u_1|u_2...u_D)du_1=1$. Once $\hat{c}_\mathbf{h}$ has been estimated, the pdf of the original forecasted variable ($x_1^\cdot$) can be estimated by applying the inverse rank transformation. The estimation of the conditional density has complexity $O(L\cdot D\cdot m)$.

\section{Estimation procedure} \label{sec:estimation-procedure}

\subsection{Multi-step prediction}
\begin{figure}[ht]
\centering
\includegraphics[width=8.8cm]{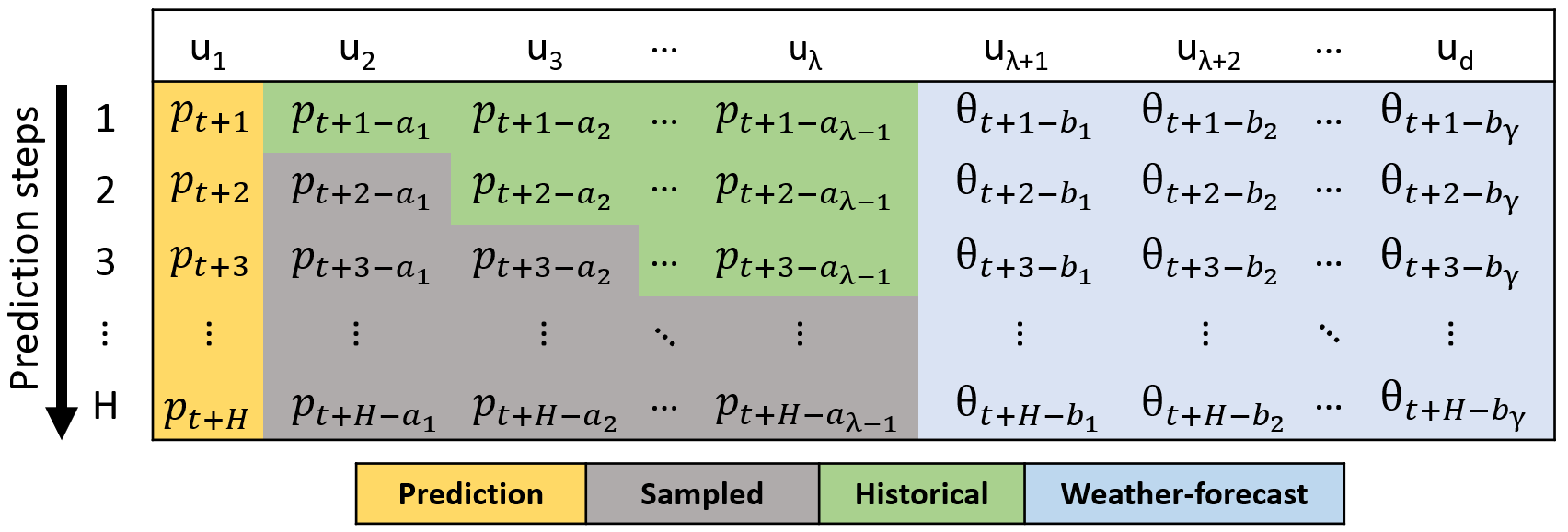}
\caption{Overview of the fixed values in the evaluation of the conditional pdf (\cref{eq:cond-density-estimation}) for multi-step prediction.}
\label{fig:prediction_steps}
\end{figure}

The conditional density estimation in \cref{eq:cond-density-estimation} requires fixing all variables $u_2,...,u_D$ to be known values, while estimating the pdf of $u_1$. Denoting $t=nT$ the starting time of the forecast, the first step of estimation, at time $t+1$ (i.e., for $j=1$), requires using historical outcomes of electricity-demand and forecasts/historical values of meteorological variables. The output of the forecast is the evaluation of the conditional pdf $\hat{c}_\mathbf{h}^1$ of $p_{t+1}$ given the available values $p_{t+1-a_1}, \ldots, p_{t+1-a_{\lambda-1}}, \theta_{t+1-b_1}, \ldots, \theta_{t+1-b_\gamma}$, in $L$ equispaced points in $[0,1]$. We denote such output $S_1\in\mathbbm{R}^L$. In the subsequent estimation-step, we want to estimate the electricity-demand at $t+2$ using $\hat{c}_\mathbf{h}^2$, which requires to fix the value for $t+1$ if $a_1=1$. If this is the case, we sample $p_{t+1}$ from the pdf obtained in the first-step. For generic forecasting steps, every unknown variable is fixed by picking a sample from the estimated pdf in a previous step. This procedure is shown in \cref{fig:prediction_steps}. In practice, as day-ahead forecasts must be produced 12-14 hours before delivery, the estimation procedure necessarily needs to forecast timesteps in the intraday before being able to forecast the day-ahead. 

\subsection{Creating scenarios and combining forecasts} \label{subsec:scenarios}
The procedure in the previous subsection creates a multi-step probabilistic forecast that we refer to as a scenario. Since we are using samples of forecasts as the conditioned variables, the procedure is run multiple times where, each time, independent samples from the estimated pdfs are drawn. I.e., we create $s$ scenarios $S^i = \{S_1^i,...,S_H^i\}, i=[1,...,s]$ where each $S_j^i \in \mathbbm{R}^L$ is an evaluation of the conditional pdf on $L$ points and $H$ is the prediction horizon. We combine the individual scenarios by summing them, each with probability $1/s$. The complete process is detailed in \cref{alg:copula}. If all the conditioned variables are available at the time of the forecast, only one scenario is necessary as there is no resampling involved.



\begin{algorithm}[ht] 
\caption{PLF with empirical copulas}\label{alg:copula}
\begin{algorithmic}
\State Initialize $s, H \gets nb\_scenarios, forecast\_horizon$
\For{$i \gets 1,s$} \Comment{\parbox[t]{.3\linewidth}{For every scenario}}
\State $p^i_k = p_k, \; k\le t$ \Comment{\parbox[t]{.3\linewidth}{Historical data up until time $t$}} \\

\For{$j\gets 1, H$} \Comment{\parbox[t]{.3\linewidth}{We forecast every time step in the forecasting horizon $t+1$ to $t+H$}} \\
\State $(u_{1},...,u_D) \gets \text{} X_k$ \Comment{\parbox[t]{.3\linewidth}{\cref{eq:data-matrix}}} \\ 
\State $\hat{c}_\mathbf{h}^j \gets \text{conditional density}$ \Comment{\parbox[t]{.3\linewidth}{\cref{eq:cond-density-estimation}}}
\State \hspace{8mm} $\text{on }p^i_{t+j-a_1}, \ldots, p^i_{t+j-a_{\lambda-1}},\theta_{t+j-b_1},\ldots,\theta_{t+j-b_\gamma}$ \\
\State $S_j^i \gets \hat{c}_\mathbf{h}^j$  \Comment{\parbox[t]{.3\linewidth}{Inverse rank transformation}} \\
\State $\text{sample } p_{t+j}^i\sim S_j^i$ \Comment{\parbox[t]{.3\linewidth}{Sample from forecast to use as conditional variable in future forecasting steps.}}
\EndFor

\EndFor
\State $S_j \gets \frac{1}{s}\sum_{i=1}^sS_j^i$ \Comment{\parbox[t]{.3\linewidth}{Averaging over every scenario.}}
\end{algorithmic}
\end{algorithm}

\subsection{Metrics} \label{subsec:metrics}
Common metrics to evaluate probabilistic forecasts include \emph{Quantile Loss} (QL), \emph{Continuous Ranked Probability Score} (CRPS), \emph{Prediction Interval Coverage Probability} (PICP) and \emph{Prediction Interval Normalized Average Width} (PINAW). The CRPS is a generalization of the Mean Absolute Error (MAE) to probabilistic forecasts and is therefore suitable to compare probabilistic and deterministic forecasts \cite{doi:10.1198/016214506000001437}. We consider $y_j$, $j\in \{1,...,H\}$ the actual outcome for a  prediction horizon $H$ and $\hat{y}_j^\alpha$, the forecasted value for every quantile $\alpha$.

The standard definition of the QL at a specific quantile $\alpha$ is:
\begin{equation}
  ql_\alpha(\hat{y}^\alpha,y) =
    \begin{cases}
       & \alpha(y-\hat{y}^\alpha), \ \ \ \ \ \ \ \ \ \ \ \ \  \hat{y}\leq y \\
       & (1-\alpha)(\hat{y}^\alpha-y), \ \ \ \ \ \  \hat{y}> y
    \end{cases}.   
\end{equation}
To summarize the performance of the full probabilistic forecast, we average the QL across 99 equidistant quantiles $0.01\leq\alpha\leq0.99$, and the forecast-horizon, $ql(\hat{y},y) = \frac{1}{99\cdot H}\sum_{j=1}^{H}\sum_{\alpha=0.01}^{0.99}ql_{\alpha}(\hat{y}_j^{\alpha},y_j) $. 

The CRPS can be approximated by averaging quantile losses across several quantiles \cite{crps-ql}, which makes it redundant to the QL metric considered here.

The PICP is defined as follows:

\begin{equation} 
    \text{PICP}(\hat{y}^\alpha,y) = \frac{1}{H} \sum_{j=1}^H\mathbbm{1}_{\hat{y}_j^{\alpha_l}\leq y_j\leq \hat{y}_j^{\alpha_u}},
\end{equation}
where $\alpha_l$ and $\alpha_u$ are the lower and upper quantiles of interest. Ideally, we would expect $\text{PICP} = \alpha_u-\alpha_l$, i.e., the observed probability of the outcome falling within a certain quantile-range equals the quantile-range. Finally, the PINAW quantifies the sharpness of the forecast, i.e., it gives a low value if the uncertainty is small. 
\begin{equation}
    \text{PINAW}_{\alpha_u,\alpha_l}(\hat{y}, y) = \frac{1}{H(\max(y)-\min(y))}\sum_{j=1}^H(\hat{y}_j^{\alpha_u}-\hat{y}_j^{\alpha_l})
\end{equation}

\subsection{Kernel bandwidth selection}  \label{sec:kernel-bandwidth-selection}
The bandwidth of the beta-kernels (e.g. \cref{eq:beta-kernel-density}) represents a hyperparameter that must be chosen. Several methods for the selection of kernel bandwidth have been studied previously, such as rule-of-thumb, plug-in-methods and cross-validation \cite{nonpareco}. The rule-of-thumb method requires the distributions to be "close" to Gaussian, and, therefore, have less appeal in a fully non-parametric setting. Plug-in-methods require knowledge of the underlying marginal distributions and are therefore challenging to implement in a real-case scenario. Finally, the cross-validation method is fully data-driven and its objective is to minimize the \emph{Integrated Square Error} (ISE) of the density estimate. That is (\cite{nonpareco}):
\begin{equation} \label{eq:ise-opt}
\begin{split}
    \min_{\mathbf{h}}\;\text{ISE} &= \min_{\mathbf{h}}{\int\left(\hat{f}_\mathbf{h}(x)-f(x)\right)^2dx} \\
    &=  \min_{\mathbf{h}}{\left(\int \hat{f}_\mathbf{h}(x)^2dx - 2\int \hat{f}_\mathbf{h}(x)f(x)dx \right)},
\end{split}
\end{equation}
where $\mathbf{h}$ is the bandwidth of the density-estimator and $f$ the true density\footnote{Note that in \cref{eq:ise-opt} we have omitted the term that doesn't depend on $\mathbf{h}$.}.

Solving the optimization problem in \cref{eq:ise-opt} requires estimating the 2nd term by the leave-one-out estimator (\cite{nonpareco}):
\begin{equation} \label{eq:ise-loo}
    \int \hat{f}_\mathbf{h}(x)f(x)dx \approx \frac{1}{m}\sum_{i=1}^m\hat{f}_{\mathbf{h},-i}(X_i),
\end{equation}
\noindent
where:
\begin{equation}
    \hat{f}_{\mathbf{h},-i}(x) = \frac{1}{m-1}\sum_{\substack{j=1 \\ j\neq i}}^mK_\mathbf{h}(x-X_j),
\end{equation}
\noindent
i.e., we estimate the density using all the data except $X_i$. Then, \cref{eq:ise-loo} calculates the average value of the density evaluated at the data-point left out.

Furthermore, in the multivariate case, the kernel bandwidths can be represented by a bandwidth-matrix $\mathbf{H}$, i.e., requiring the optimization of $\frac{D(D+1)}{2}$ parameters. Even by restricting $\mathbf{H}$ to have only diagonal elements (i.e., one bandwidth per dimension, denoted $\mathbf{h}\in\mathbbm{R}^D$) results in a very challenging problem to solve considering its nonlinear nature. The estimation of a joint density using the beta kernel is also computationally heavy and therefore not practically implementable. Instead, we resort to the conditional density-formulation presented in \cref{subsec:conditional-density-estimation}. The ISE can therefore be reformulated as follows:

\begin{equation} \label{eq:conditional-ISE}
\begin{split}
    \min_{\mathbf{h}}\;\text{ISE} &= \int_0^1 \left( \hat{c}(u_1|u_2,...,u_D;\mathbf{h}) \right)^2 du_1 \\
     & - \frac{2}{m}\sum_{i=1}^m \hat{c}_{-i}(u_1|u_2=r(x_2^i),...,u_D=r(x_D^i);\mathbf{h}),  
\end{split}
\end{equation}
\noindent
where the leave-one-out estimator takes the form:
\begin{equation} 
    \hat{c}_{-i}(u_1|u_2...u_D) \propto \sum_{\substack{j=1 \\ j\neq i}}^m\prod_{k=1}^DK\left(r(x_{k}^{j}), \frac{u_k}{h_k}+1, \frac{1-u_k}{h_k}+1 \right).
\end{equation}

The optimal bandwidths in \cref{eq:conditional-ISE} are, generally, only optimal for a fixed value of the conditioned variables and not for any possible value. However, as hypothesized in \cref{subsec:hypothesis}, electricity demand have strong serial dependence and we therefore expect conditioned variables between different days to remain ``close''. It is also interesting to note that the optimization of \cref{eq:conditional-ISE} does not involve the realization of the forecasted variable. Therefore, the optimization can be performed at the time of the forecast, as opposed to other optimization-techniques such as minimizing the QL (see below).

The rule-of-thumb bandwidth selection for multivariate distributions can be formulated as follows \cite{https://doi.org/10.1002/bimj.4710300745}:
\begin{equation} \label{eq:rule-of-thumb}
    h_j = \left( \frac{4}{D+2}\right)^{\frac{1}{d+4}}\frac{1}{m^{\frac{1}{D+4}}}\sigma_j,
\end{equation}
for $j\in\{1,...,D\}$, $D$ the number of dimensions, $m$ the number of samples and $\sigma_j$ the standard deviation of the $j$-th dimension. Since the data has been rank-transformed, the standard deviation of every variable is fixed and approximately equal to 0.29. Therefore, the rule-of-thumb bandwidth is deterministic and only a function of the number of samples and dimensions of the copula.

Although optimizing the bandwidths using objectives involving the shape of the underlying joint/conditional pdf ensures optimal kernel density estimation, it does not necessarily produce the best forecasts. Additionally, if the number of dimensions of the copula is large, these methods suffer from the ``curse of dimensionality'' since the distance between every sample tends to converge. This results in very small ISE-optimal bandwidths or the need of very large dataset of historical measurements, and therefore the selection of variables (dimensions) of the copula, becomes restrictive. If instead, one formulates an objective using a metric for the performance of the forecast, such as QL or CRPS, this limitation is avoided. In this case, the number of dimensions of the copula can be arbitrarily large, at the expense of being an ill-posed problem. We propose the following objective:

\begin{equation} \label{eq:pl-opt}
\begin{split}
    \min_{\mathbf{h}}\;\text{QL} &= \min_{\mathbf{h}}{ql(\hat{y},y)} \\
    &=  \min_{\mathbf{h}}{\sum_{j=1}^{H}\sum_{\alpha=0.01}^{0.99}ql_{\alpha}(\hat{y}_j^{\alpha},y_j)},
\end{split}
\end{equation}
where $\hat{y}_j^{\alpha}$ is the $\alpha$-quantile of the $j$-th forecasting step, $0.01\leq \alpha\geq 0.99$, $H$ is the number of time steps in one day and $ql_\alpha$ as defined in \cref{subsec:metrics}. Note that this objective depends on the realization of $y$ and must therefore be run on historic data, for example, the same day one week before. The formulation in \cref{eq:pl-opt} results in one set of bandwidths, $\mathbf{h}$, for every weekday.

\section{Case-study} \label{sec:case-study}
\subsection{Data}
\begin{figure}[ht]
\centering
\includegraphics[width=8cm]{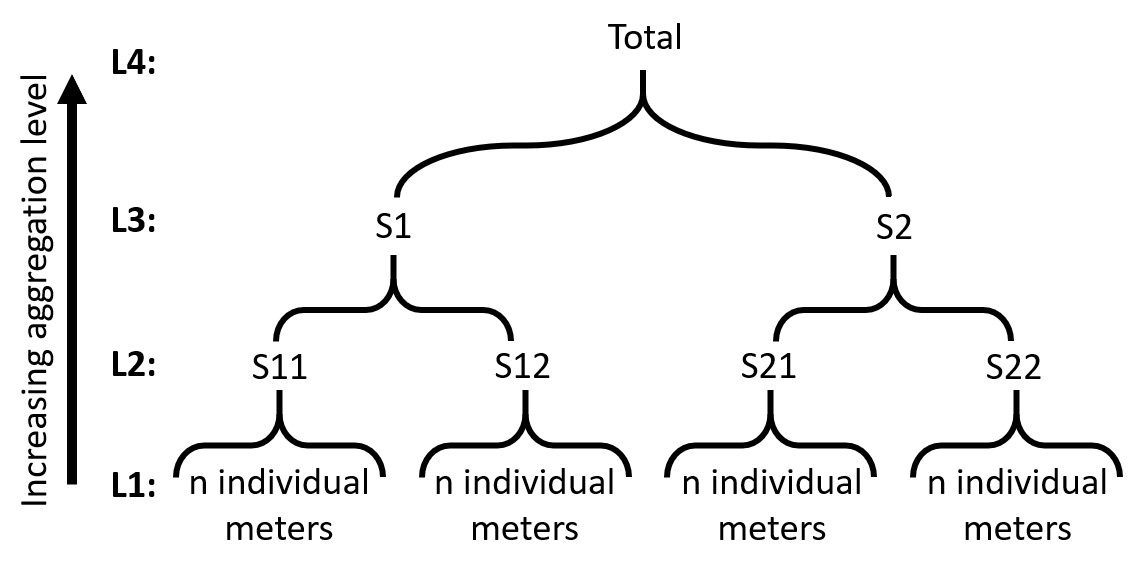}
\caption{Aggregation-structure of the data used to evaluate the model.}
\label{fig:data_structure}
\end{figure}

The availability of open data time series of electricity consumption at a low aggregated level is limited. We use the data in \cite{NESPOLI2020106755}, collected from power meters installed in secondary substations and low-voltage cabinets in Rolle (Switzerland). The aggregated peak-demand in the measurement-period was 1134 kW, the mean 655 kW with a standard deviation of 141 kW. This comprises a total of 24 measurement-points (end-users) where the mean demand varies between 9.5 kW and 52 kW. The data also includes historical meteorological forecast data from a commercial provider. The structure of the data can be seen in \cref{fig:data_structure}, where we have added the nomenclature L1-4 to highlight the different aggregation levels. L1 corresponds to the end-users measurements and L2-4 corresponds to an increasing level of aggregation. The individual time series was gathered between January 2018 and January 2019 with 10-minutes spacing. We down-sample the data to 15-minutes spacing to reduce the number of forecasting steps and better represent electricity-markets, which currently have a minimum contract-size of 15 minutes \cite{epexspot}.

\subsection{Bandwidth-optimization}
We investigate the forecast performance using the three different bandwidth-optimization routines outlined in \cref{sec:kernel-bandwidth-selection} (\cref{eq:ise-opt,eq:pl-opt,eq:rule-of-thumb}). For the ISE and QL-optimization we use the L-BFGS-B algorithm implemented in \cite{2020SciPy-NMeth} with initial guess $h_i = 0.01, i=1,...,D$. In the case of ISE-optimization, we observed that the optimal bandwidths are always the same for different initializations while, for the QL-optimization, it varies, suggesting there are local minima\footnote{Assessing the convexity of the ISE- and QL-optimization is challenging and outside the scope of this paper. This specific problem will be investigated in future research by the Authors.}. As mentioned in \cref{sec:kernel-bandwidth-selection} the ISE-optimization can be performed at the time of forecasting and does not require a separate training dataset. The QL-optimization is run on days between 10-16 December 2018 to obtain one $\mathbf{h}$-vector per day of the week.

\subsection{Selection of lags and pre-clustering}
The number of lagged variables is considered a design parameter and depends on the observed temporal dependence of the data. When optimizing bandwidths using ISE as objective, the number of lags was restricted to the temperature at lag $0$ and historical demand at lag $672$ which is equivalent to the same time step one week before. This results in an empirical copula with 3 dimensions (the dimension for the forecasted time step and one dimension for every conditional variable). When using QL and rule-of-thumb as the objective in the bandwidth-optimization, we investigated two models with different lags (see \cref{tab:lags}). The models using lagged demand not available at the time of the forecast were run multiple times to create 50 scenarios which then was combined as explained in \cref{subsec:scenarios}. The temperature-lag is 0 because it corresponds to the forecasted value for the actual time-step. In principle, the lags of forecasted variables can also be negative, i.e., forecasts for future time-steps. This can be useful if the system contains loads that adapts to forecasts, such as smart heating systems. 

\begin{table}[ht!]
\centering
\caption{Selected lags for QL-optimized and rule-of-thumb optimized models. As mentioned, the sampling time is 15 minutes.}
\renewcommand{\arraystretch}{1.3}
\begin{tabular}{@{}lll@{}} \toprule
& Electricity demand lags  & Temperature forecast \\ 
\midrule
ISE-optimized & \{672\} & \{0\}\\
QL-optimized 1 & \{672\} & \{0\}\\
QL-optimized 2 & \{1, 12, 24, 96, 672\} & \{0\} \\
Rule-of-thumb-optimized 1 & \{672\} & \{0\}\\
Rule-of-thumb-optimized 2 & \{1, 12, 24, 96, 672\} & \{0\} \\
\bottomrule
\end{tabular}\label{tab:lags}
\end{table}

For both the QR-model (see \cref{subsec:benchmark}) and our model, the data is pre-clustered into working days and holidays since these have different distributions of electricity demand. The clustering is done using the Python package \textit{Workalendar}, taking into account all local holidays at the location of the power meters \cite{workalendar}.  Clustering creates discontinuities in the data, which influences the performance of the forecasts. In our model, the lagged values are calculated before clustering to preserve the temporal dependence. 

\subsection{Forecast-horizon for market participation}
The clearance of day-ahead wholesale electricity markets happens around 12 hours before beginning of delivery\footnote{In Switzerland, day-ahead auction clearance happens at 11am D-1, while most other European countries clear at noon. \cite{epexspot}}. Therefore, our model is fed with data until 10am and then provides a forecast for the next day, to adhere to the practical needs of DSOs and other market participants.

\subsection{Benchmark against Quantile Regression} \label{subsec:benchmark}
As previously mentioned, the proposed method can be useful to forecast at different aggregation-levels. Then, individual forecasts can be aggregated to produce an overall forecast. A natural question is therefore at which level to forecast, to obtain the most accurate overall forecasts. This question is treated in \cref{sec:app1}. Aggregating forecasts by QR is not the goal of this paper and we therefore resort to performing the benchmark on aggregation level L4 (total electricity demand). The data pre-processing for the QR model is equivalent to the one previously explained. For a fair comparison, the considered lags are equivalent to the lags in the ISE-optimized empirical copula model. In the QR framework, we seek to find the parameters of the following equation:
\begin{equation}
    p_t^{\alpha} = \theta_{0}^\alpha  + \theta_1^\alpha p_{t-672} + \theta_2^\alpha T_{t},
\end{equation}
where $p_t^{\alpha}$ is the $\alpha$-quantile of the forecasted demand for time $t$, $p_t, p_{t-672}$ is the power at time $t$ and 692 steps before (same day, previous week) and $T_t$ is the temperature-forecast for time $t$. The weights $\pmb{\theta}^\alpha = [\theta_0^\alpha,\theta_1^\alpha,\theta_2^\alpha]$ are unique for every quantile and cluster and they are found using \cite{seabold2010statsmodels} to estimate 99 quantiles of the forecasted power, equally spaced between 0.01 and 0.99.

\begin{table}[ht]
\centering
\caption{Results from 7 forecasted days at aggregation-level L4 (i.e. aggregation of all the data). The 5 EC methods using different bandwidth-optimization routines are compared against the quantile regression model. The quantile loss along with the PICP and PINAW at 5-95\% and 10-90\% prediction interval is reported.}
\renewcommand{\arraystretch}{1.3}
\begin{tabular}{@{}llllll@{}} \toprule
& \multirow{2}{*}{QL}  & \multicolumn{2}{c}{PICP} & \multicolumn{2}{c}{PINAW} \\ \cmidrule(r){3-4} \cmidrule(r){5-6}
 &  & 5-95 & 10-90 & 5-95 & 10-90\\ \midrule

EC ISE-optimized    &   9.874 &   0.946 &       0.902 &        0.305 &        0.239 \\
EC QL-optimized 1    &   9.915 &       0.951 &       0.888 &        0.299 &        0.23 \\
EC QL-optimized 2    &   11.954 &       0.847 &       0.778 &        0.25 &        0.196 \\
EC Rule-of-thumb optimized 1    &   10.641 &       0.982 &       0.958 &        0.389 &        0.315 \\
EC Rule-of-thumb optimized 2    &   11.915 &       0.978 &       0.93 &        0.379 &        0.31 \\
Quantile regression &  12.023 &       0.976 &       0.924 &        0.424 &        0.308 \\

\bottomrule
\end{tabular}\label{tab:res-all}
\end{table}

The models are run for 7 individual days between 13 January to 19 January, 2019 to include both weekend and weekday-dynamics, as well as the transition between them. The results are presented in \cref{tab:res-all}. Overall, we observe that the ISE-optimized empirical copula has an 18\% decrease of QL compared to the QR model. The prediction coverage (PICP) is satisfactory for both models, while the ISE-optimized EC has much narrower prediction-intervals as measured by the PINAW. The alternative EC-models based on QL-optimization and rule-of-thumb optimization shows less improvement compared to quantile regression, when measured by QL, however, the QL-optimized ECs produce sharper forecasts (i.e. smaller PINAW). 

\begin{figure*}
     \centering
     \begin{subfigure}[b]{0.48\textwidth}
         \centering
         \includegraphics[width=\textwidth]{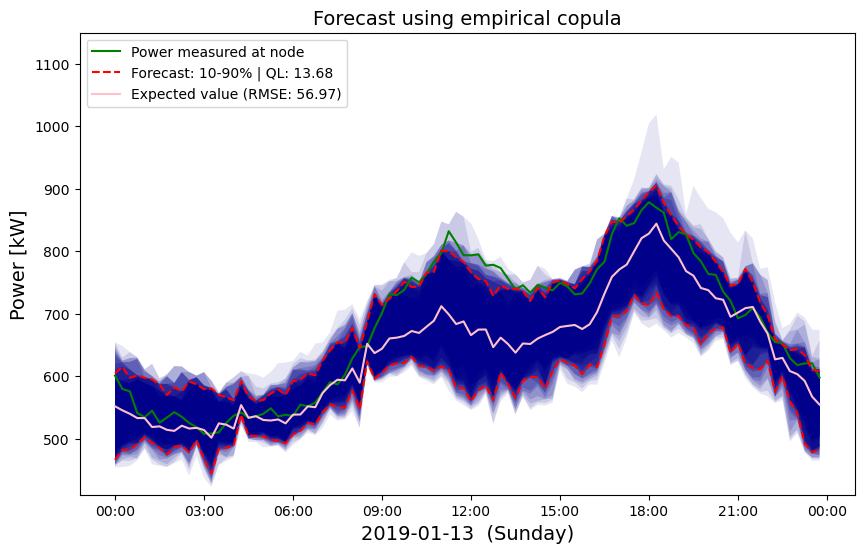}
         \caption{Empirical copula.}
         \label{fig:ec-weekend}
     \end{subfigure}
     \hfill
     \begin{subfigure}[b]{0.48\textwidth}
         \centering
         \includegraphics[width=\textwidth]{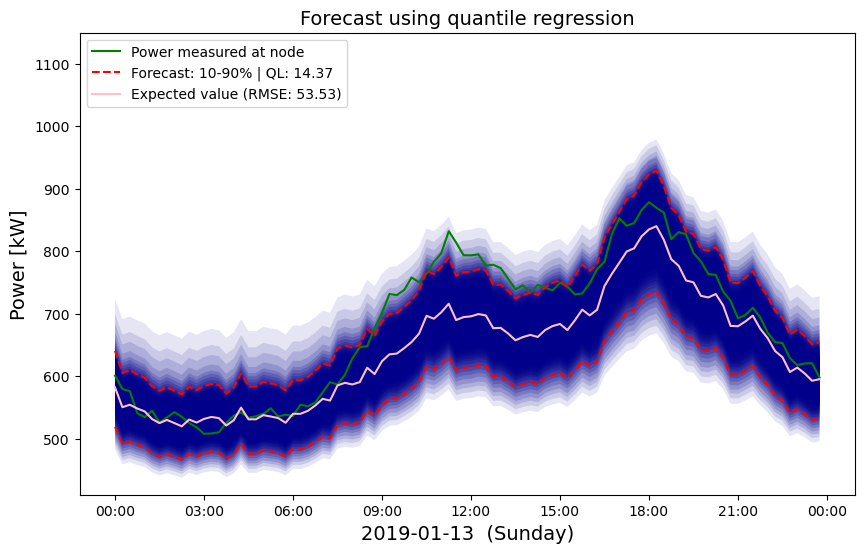}
         \caption{Quantile regression.}
         \label{fig:qr-weekend}
     \end{subfigure}
        \caption{Example of day-ahead forecast for a weekend-day. Left: empirical copula. Right: quantile regression. The green and red curve show the actual outcome and the 0.5-quantile of the forecast, respectively. The pink curve shows the expected value. The blue shading shows the different quantiles. The color-gradient goes from darkest blue at the 0.5-quantile to the lightest blue at the 0.01 and 0.99-quantiles.}
        \label{fig:weekend}
\end{figure*}

\begin{figure*}
     \centering
     \begin{subfigure}[b]{0.48\textwidth}
         \centering
         \includegraphics[width=\textwidth]{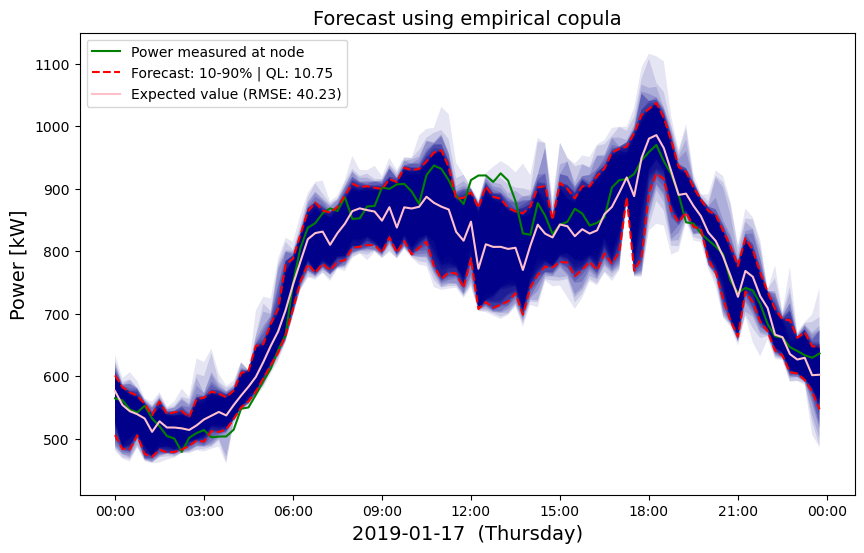}
         \caption{Empirical copula.}
         \label{fig:ec-weekday}
     \end{subfigure}
     \hfill
     \begin{subfigure}[b]{0.48\textwidth}
         \centering
         \includegraphics[width=\textwidth]{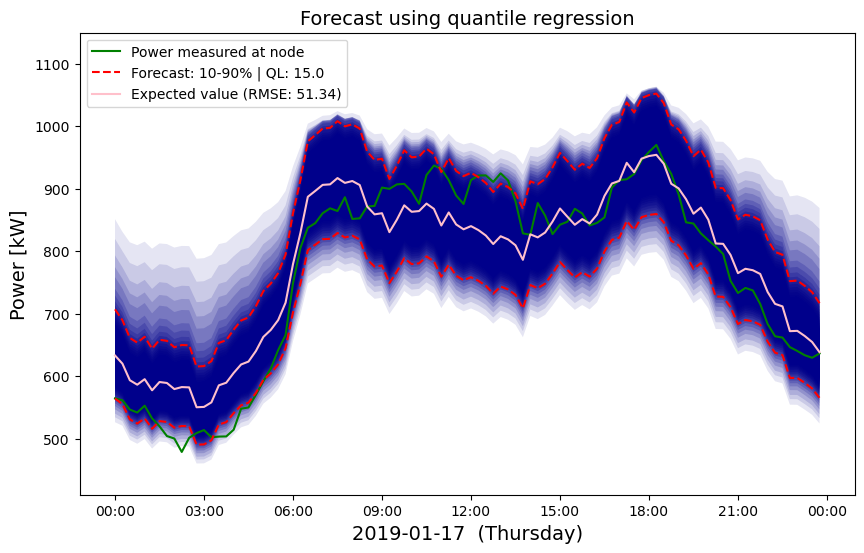}
         \caption{Quantile regression.}
         \label{fig:qr-weekday}
     \end{subfigure}
        \caption{Example of day-ahead forecast for a working-day. Left: empirical copula. Right: quantile regression. It is observed that our model provides narrower confidence intervals, especially in the morning ramp-up and the evening ramp-down. Confidence intervals are also more variable over the prediction horizon, highlighting the non-linear, data-driven nature of the model.}
        \label{fig:weekday}
\end{figure*}

As an example, the forecasts for two individual days are shown in \cref{fig:weekend,fig:weekday}, one for a weekend-day (Sunday) and one for a working-day (Thursday). The left plot shows the EC while the right plot shows the QR. We observe that our model seems to better forecast the actual outcome and at times, provide sharper confidence intervals. In particular, it provides sharper confidence intervals during the morning ramp-up and the evening ramp-down on weekdays. 

\section{Conclusion} \label{section:discussion-conclusion} 

In this paper we have presented a probabilistic forecasting model based on empirical copulas. The model is fully data-driven and makes no assumptions on the distributions of variables. It is highly adaptable and naturally supports ordinal variables of any kind. The model was applied in the context of electricity load forecasting of distribution-systems at different aggregation levels, and included weather-forecasts of temperature. Low-granularity electricity load timeseries are stochastic in nature and assumptions such as linearity and normally distributed variables are not always valid. The model can provide meaningful forecasts at low-granularity level and overall, we observe 18\% reduction of \emph{Quantile Loss} (QL), compared to a \emph{Quantile Regression} (QR) model. The non-parametric approach allows the confidence intervals to be asymmetric, better representing the actual uncertainty in the process, as opposed to parametric methods that (usually) relies on symmetric probability-distributions. The method can be extended to include any numerical variable, such as solar irradiance and wind speed/direction, allowing forecasts of \emph{Photo Voltaics} (PV) and wind power.

\appendix
\section{Optimal aggregation-level}
\label{sec:app1}
We investigate the optimal aggregation-level that produces the best overall forecasts. Distribution-grids are usually radial with power-measurements at different aggregation-levels (Ref. \cref{fig:data_structure}). It is therefore possible to forecast at different aggregation-levels depending on the objective of the forecast. In this Appendix, the objective is to forecast the overall consumption of a urban district (Total), by either aggregating loads and then forecast, or aggregating forecasts of individual consumers meters. The aggregation of probabilistic forecasts is done by sampling independently from every individual forecast and then summing the samples.

\begin{table}[ht]
\centering
\caption{Forecast at different aggregation-level for the chosen lags and ISE-optimal kernel bandwidths.}
\renewcommand{\arraystretch}{1.3}
\begin{tabular}{@{}p{1cm}lp{2cm}p{2cm}p{2cm}p{2cm}@{}} \toprule
& \multirow{2}{*}{QL}  & \multicolumn{2}{c}{PICP} & \multicolumn{2}{c}{PINAW} \\ \cmidrule(r){3-4} \cmidrule(r){5-6}
 &   & 5-95 & 10-90 & 5-95 & 10-90\\ \midrule

L1 &  10.179 &       0.771 &       0.658 &        0.147 &        0.114 \\
L2 &  9.832  &       0.888 &       0.789 &        0.2 &        0.157 \\
L3 &  9.86 &       0.924 &       0.869 &        0.251 &        0.197 \\
L4 &  9.874 &   0.946 &       0.902 &        0.305 &        0.239 \\

\bottomrule
\end{tabular}\label{tab:aggregation-level}
\end{table}

In \cref{tab:aggregation-level} the results from forecasting the aggregated load are shown. Every row corresponds to an aggregation-level where first, individual forecasts are performed, followed by aggregating the individual forecasts. L4 corresponds to the total aggregated load and is thus directly forecasted. All metrics are calculated as the average over 7 forecasted days between 13. January and 19. January, 2019. As can be seen, in this specific example, the optimal (w.r.t. QL) aggregation-level is L2. This indicates that it is beneficial to forecast $S11$, $S12$, $S21$ and $S22$ (ref \cref{fig:data_structure}) individually, then aggregating them into the final (total) forecast. 

\bibliographystyle{unsrt}
\bibliography{Bibliography}






\end{document}